\documentclass[aps,pra,10pt,floatfix,twocolumn,superscriptaddress]{revtex4-2}

\usepackage{graphicx,bm,dsfont,amsmath,amsthm,amssymb,mathrsfs,hyperref,mdframed,bbold,hyperref,xcolor,soul}
\usepackage[utf8]{inputenc}
\usepackage[toc,page]{appendix}
\usepackage{orcidlink}

\newcommand{\ket}[1]{\left| #1 \right\rangle}

\newcommand{\ketbra}[2]{\left|#1\right\rangle\hskip-1mm\left\langle #2\right|}

\begin{document}

\title{Experimental Tests of Invariant Set Theory}

\author{Jonte R. Hance\,\orcidlink{0000-0001-8587-7618}}
\email{jonte.hance@newcastle.ac.uk}
\affiliation{School of Computing, Newcastle University, 1 Science Square, Newcastle upon Tyne, NE4 5TG, UK}
\affiliation{Quantum Engineering Technology Laboratories, Department of Electrical and Electronic Engineering, University of Bristol, Woodland Road, Bristol, BS8 1US, UK}
\author{Tim N. Palmer}
\affiliation{Department of Physics, University of Oxford, Oxford, UK}
\author{John Rarity}
\affiliation{Quantum Engineering Technology Laboratories, Department of Electrical and Electronic Engineering, University of Bristol, Woodland Road, Bristol, BS8 1US, UK}

\date{\today}

\begin{abstract}
We identify points of difference between Invariant Set Theory and standard quantum theory, and show that these lead to noticeable differences in predictions between the two theories. We design a number of experiments to test which of these predictions corresponds to our world. If these experiments were undertaken, they would allow us to investigate whether standard quantum theory or invariant set theory best describes reality. These tests can also be deployed on theories sharing similar properties (e.g., Penrose's gravitational collapse theory).
\end{abstract}

\maketitle
\section{Introduction}
For all the successes of modern physics over the last century-and-a-half, we have been left with two apparently incompatible branches - the nonlinear and deterministic General Relativity, and the linear but indeterminate quantum theory. For us to have a Theory of Everything, that describes all observed physical phenomena, we need a way to unite these, so we can describe physical phenomena at any scale. However, due to their differing takes on the determinacy of the universe, this has so far proved difficult.

Invariant Set Theory (IST) attempts to unify these two disparate branches by using insight from Chaos Theory to create a fully local and determinate model of quantum phenomena \cite{Palmer1995Spin,Palmer2009ISP,palmer2016invariantsettheory,Palmer2020Discretization}. It does this by assuming that the universe is a determinate dynamical system evolving precisely on a fractal invariant set in state space. The natural metric to describe distances on a fractal set is the $p$-adic metric {(a fractal metric, with different properties to the Euclidean metric with which define spatial distance)}.

{In IST,} this $p$-adic metric replaces the standard Euclidean metric of distance between states in state space. A consequence of this switch is that putative counterfactual states which lie in the fractal gaps of the invariant set are considered distant from states which do lie on the invariant set, even though from a Euclidean perspective such distances may appear small. {In IST, this is used as an argument for these states being counterfactually restricted \mbox{\cite{Hance2024CFRestrict}}---in the same way as spaces between points on a ``snap-to'' grid on a computer are inaccessible, these $p$-adically distant states are inaccessible in IST.} Given {which points are on} any given fractal attractor {is formally uncomputable}, we cannot in advance distinguish states allowed and disallowed by this metric. Hence, in IST, quantum-scale phenomena appear random despite being deterministic. 

$p$-adic numbers form a back-bone of modern number theory {\mbox{\cite{Woodcock1998padic}}} and as such provide {way for us to use sophisticated tools from finite number theory to} describe quantum physics. An example of this is how IST explains complementarity, a concept underpinning the uncertainty principle in quantum mechanics. In IST, complementarity is an emergent phenomenon arising from Niven's theorem, a number-theoretic property of trigonometric functions. {The theorem states that} $\cos \phi$ is not a rational number when $\exp {i \phi}$ is a primitive $p$th root of unity (i.e., when $\phi$ is not 0 or some integer multiple of $\pi$). Notably, the complex Hilbert Space of standard quantum mechanics {seems to arise} as a singular limit of invariant set theory when $p$ is set equal to infinity. 

However, despite showing how key examples of quantum phenomena (like the sequential Stern-Gerlach effect, and Bell inequality violation, {in \mbox{\cite{Palmer2020Discretization}}}) can be described deterministically, the theory deviates from standard quantum physics in some of its predictions---mainly in ways which stem from the $p$-adic metric being finite. {While there have been some attempts to challenge IST mathematically \mbox{\cite{Sen2021Analysis}}, these empirical differences have not yet been properly considered.} In this paper, we give these key points of deviation, and investigate the extent to which these could be used to experimentally test the theory.

\section{Invariant Set Theory}
IST is an extension of quantum mechanics based on the assumption that the universe is a deterministically-evolving system, where the allowable set of states in state-space is a fractal. This fractal set, $I_U$, is such that if a state lies in the set, no time evolution of that state takes it out of the set; conversely, if a state is not in the set, no time evolution can bring it into the set. The measure of this invariant set, $\tilde \mu$, is non-trivial such that the model can violate Bell inequalities \cite{Hossenfelder2020Rethinking,Hossenfelder2020Perplexed,Hance2021Ensemble}---the model is supermeasured \cite{Hance2022Supermeasured}.

Given the allowed set of states is a fractal, it has gaps. (If anything, given the set has measure zero compared to state space, it is mostly gaps.) States in the gaps are counterfactual states which are mathematically possible, but are physically unrealisable---they are counterfactually restricted by IST \cite{Hance2024CFRestrict}. By Euclidean distance, the restricted states seem arbitrarily close to allowed states in state space. Tiny perturbations (from the point of view of this Euclidean metric) generically take allowed states to those that are counterfactually restricted by the model. However, these perturbations are necessarily massive when considered using a $p$-adic notion of distance, which is the natural notion to use on a fractal geometry. Therefore, even though this restriction may seem arbitrary and fine-tuned when considered with a Euclidean metric, it is arguably well-motivated when considered using a fractal set of allowed states, as IST posits. In fact, many areas of arithmetic dynamics already model dynamical systems using $p$-adic numbers (see e.g., \cite{Woodcock1998padic}). 

The Lorenz model provides an illustrative example of the sort of restriction we see in IST. No matter where in state space we initialise the three ordinary differential equations which make up this model, after an infinite amount of time, the trajectories given as solutions of these equations fall onto the fractal Lorenz attractor. One difference though between this illustrative model and IST is that IST proposes that laws of physics are not based on differential equations, but instead on geometric equations describing the attractor itself. Point on state space which do not lie on the attractor are not physically consistent with these laws, meaning these points are assigned probability zero. In IST, the reason we cannot simply deterministically compute the trajectories of these points, and so describe all systems in a classical deterministic manner, is that we cannot know \textit{a priori} whether a given point in state space lies on this attractor: the geometric properties of these sorts of fractal structures (e.g., the Lorenz attractor) are formally incomputable \mbox{\cite{blum1998complexity,Dube1993Fractal}}.

{There is not yet} a written dynamical law for IST. However, this is common for quantum foundational models, as we often only care about transition amplitudes between initial and final times. Further, these amplitudes are often between simple qubit observables (e.g., spin or polarisation states), meaning these models often don't give or require a space-time evolution law either. Spekkens's Toy Model is an example of such a model \cite{Spekkens2007EpistTot}, having no dynamical evolution equation but still proving useful for considering foundational questions. Similarly to this model, IST in its current form exists to see what interesting insights we can get from considering incomputable counterfactual restrictions on state space.

Due to its fractal structure, in IST we can expand the state of a system in a detector-eigenstate basis $|A_j \rangle$ as
\begin{equation}\label{Eq:Expansion}
|\psi\rangle = a_1 |A_1\rangle + a_2|A_2\rangle \ldots + a_J |A_J\rangle~,
\end{equation}
where $\sum_j a_j a^*_j =1$.
However, unlike standard quantum mechanics, these complex amplitudes $a_j$ must be such that, if we write $a_j$ in polar form $a_j=R_j e^{i \phi_j}$,
\begin{equation}
\label{Eq:finite}
R^2_j=m_j/p; \ \ \ 
\phi_j = 2\pi n_j/p~,
\end{equation}
where $m_j,n_j,p \in {\mathbb N}_0$, $m_j,n_j < p$. Effectively, these amplitudes must have both discretised magnitude and phase. This discretisation gives measure zero to any states whose coefficients in Eq.~\ref{Eq:Expansion} do not obey the conditions in Eq.~\ref{Eq:finite}, meaning distributions over the set of allowed states can violate (Bell-)Statistical Independence, and so be used to violate Bell inequalities, even when the distributions themselves contain no information about the detector settings.

If we take $p \rightarrow \infty$, the set of such ``rational'' Hilbert states becomes dense over the projection of the Hilbert space we get from standard quantum mechanics---we can always find an allowed state ``close enough'' to any restricted state which we may want to prepare or investigate. Therefore, if we make a model where $p$ is large enough, we can make IST as experimentally indistinguishable from quantum theory as we want. This makes it difficult to devise an experimental test discriminating between IST and standard quantum mechanics: any test we develop must rely on $p$ being a finite number.

However, no matter how large $p$ is, the state-space of this theory will continue to have gaps---the limit $p \rightarrow \infty$ is singular, meaning quantum mechanics does not correspond to IST in the large $p$ limit. Specifically, no matter how large $p$ is, if a state does not obey the rationality conditions given above, it is counterfactually restricted. 

This is the mechanism IST uses to explain why we cannot simultaneously measure conjugate variables in quantum mechanics with certainty. While in quantum mechanics, this is a consequence of having non-commuting operators acting on a Hilbert-space, in IST this arises due to the geometric structure of the invariant set and associated fractal measure. The incomplete algebraic structure of the set of allowed amplitudes reflects the ``gappy'' geometric structure of this fractal set. For example, by combining the amplitudes above with Niven's theorem, we can see that superpositions of two states which are allowed by IST are generically not also allowed. 

Obviously, thinking about rational numbers and restrictions on which states can mutually exist does not explain all results of quantum mechanics---for that, we would need other things, such as a dynamical law. However it does present a mathematical mechanism for us being unable to simultaneously measure certain combinations of properties of quantum systems.

In the following we do not directly use the fractal structure of an invariant set. Invariant sets generically being fractals motivates us to consider a finite discretisation of Hilbert space, where certain combinations of states do not exist.

{Instead, we use the fact that we can build a representation of a qubit obeying the discretisation conditions given in Eq.~\mbox{\ref{Eq:finite}} using a bit-string of length $p$. To explore empirical differences between IST and standard quantum mechanics, we following \mbox{\cite{Palmer2020Discretization}} to do so. Note though that we only describe the elements of this which are necessary to motivate the experimental differences we discuss below---if you are more interested in how such a bit-string model of qubits in IST can be used to demonstrate complementarity, or to violate a Bell inequality, we strongly suggest you read through \mbox{\cite{Palmer2020Discretization}}. In this bit-string, amplitude (treated as rotation from the top pole of the Bloch sphere in our basis of choice) can be represented by proportion of bits taking one value; and phase (treated as rotation around the circumference of the Bloch sphere) can be represented as cyclic permutation of this bit-string from the scenario where it is, in-order, first all 0-bits, then all 1-bits---i.e., if the value of the $i^th$ bit in the bit string is $a_i$, phase $\phi = 2\pi n/p$ causes cyclic permutation:}
\begin{equation}
    \zeta^{n = \phi p/2\pi}\{a_1, a_2,...a_p\} = \{a_{n+1}, a_{n+2},...a_p, a_1,a_2,...a_n\}
\end{equation}
{This makes sense, given, at either pole of the Bloch sphere (relative to this representation), the bit-string is either all 0s or all 1s, so a cyclic permutation of the bits in the string does nothing to the bit-string. The various unitary operations which can be done on a qubit can all be represented as rotations on the Bloch sphere, so can all be represented as changes to the values of bits in the bit-string (increasing or decreasing the relative proportion of 0s and 1s to change the latitude of the state on the Bloch sphere), to the cyclic permutation of the position of the bits in the string relative to the ``phaseless'' bit-string}
\begin{equation}
    \{0_1, 0_2,...0_{p-m}, 1_{p-m+1},...1_p\}
\end{equation}
{where, for an arbitrary qubit $\ket{\psi} = \alpha\ket{0}+\beta\ket{1}$, $|\beta|^2 = m/p$.}

{This shows how IST uses bit-strings to represent a single qubit, but what about multiple, potentially-entangled qubits?}

{To do this, for $J$ qubits, IST says we need to take the Cartesian product of $J$ bit-strings, each of length $p$. Where the ``phaseless'' form of the first bit-string is partitioned as $p-m_1$ 0s, then $m_1$ 1s, the ``phaseless'' form of the second bit-string is here partitioned $p-m_2$ 0s, then $m_2-m_1$ 1s, then $m_3$ 0s, then $m_1-m_3$ 1s. This means the first sub-string of 0s then 1s of the second bit-string is the same length as the chain of 0s of the first bit-string ($p-m_1$), and the second sub-string of the second bit-string is the same length as the chain of 1s in the first bit-string ($m_1)$. When there is exactly the same ratio of 0s to 1s in the first sub-string as there is in the second sub-string, then obviously the state of the preceding string (or rather, preceding qubit) has no effect on the state of that qubit---therefore, the two qubits are unentangled. However, when the state is maximally affected (as in, there are say no 0s in the first sub-string, and no 1s in the second sub-string), then this is effectively a maximal entanglement. Beyond this phaseless case, we also now have three different cyclic permutation operations---one on the first-bit-string, as before, but also one on the first sub-string of the second bit-string, and one on the second sub-string of the second bit-string. As well as in correlation between weightings of 0s and 1s, entanglement can also manifest in correlations imposed by these cyclic permutations (i.e., there being difference between the number of times the cyclic permutation operator is applied to the first sub-string of the second bit-string, and the number of times it is applied to the second sub-string). For the two qubits, this gives us 6 free integer parameters. Given quantum mechanics, for two qubits, requires a state space $\mathcal{S}^6$, this should (up to a certain resolution) allow us to represent two entangled qubits. Note though that, due to the partitioning of the second qubit into two sub-strings, we effectively half the resolution which we had previously for a single qubit.

This subdivision (in the phaseless case) goes similarly for a third bit-string (with has four sub-strings, two for each of the two sub-strings of the second bit-string), and for the fourth bit-string (which has 8 sub-strings, two for each of the four sub-strings of the second bit-string), and so on, for as many qubits as we wish to add to the system. Entanglement between any two of the qubits can manifest through differences in relative weighting of 0s and 1s in certain sub-strings, differences in number of times cyclic permutation is applied in certain substrings, or a combination of these two effects. However, the more qubits are added to the system, the lower the resolution with which the systems are able to be represented within the $p$-length bit-strings, and so the smaller the amount of entanglement-induced correlation can be represented on the system. In order for us to be able to use Niven's Theorem to reproduce complementarity for a single qubit, \mbox{\cite{Palmer2020Discretization}} shows that we need a bit string of length $p=2^M$, where $M\geq2$. Similarly, given the progressive partitioning of bit-strings into sub-strings as we add potentially-correlated qubits to the system, for our system of $J$ qubits to behave in any way quantum-mechanically together, we need $M\geq 2J$, and so $p\geq 2^{2J}$. Put another way, IST presents a limit of $J = \log_2(p)/2$ qubits which can be entangled together.}

\section{Entanglement Limits}

\begin{figure}
    \centering
    \includegraphics[width=\linewidth]{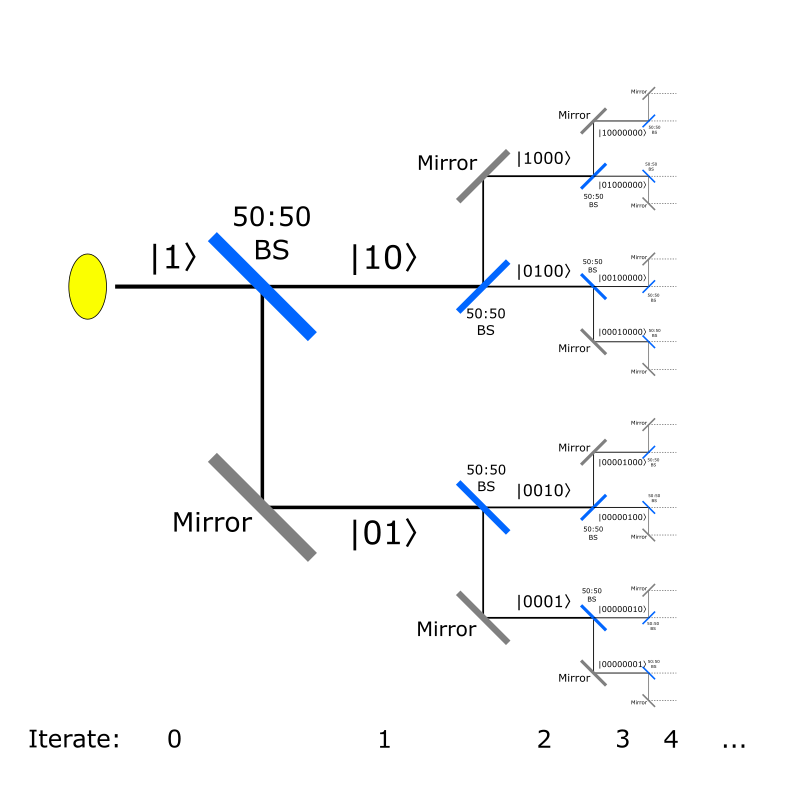}
    \caption{The first 4 iterates of a set-up to create the entangled state $\ket{W^J}$ (as given in Eq.~\ref{EqW}), where $J=2^I$ at the $I^{th}$ iterate. The diagonal blue lines are 50:50 beamsplitters, the diagonal grey lines mirrors, the yellow oval a single-photon source, and the black lines the possible paths of the photon. Given this maximally entangles $2^I$ qubits, IST predicts entanglement generated by an experiment like this should begin to fail after $I=\log_2\log_2p$ iterations, where the two spherical dimensions of the Bloch sphere are each $p$-discrete. We can test whether this entanglement holds or fails by putting mirrors at the ends of each path - if the photon returns with 100$\%$ probability to the input port, it was maximally entangled; if each beamsplitter splits it evenly, such that it only returns $2^{-I}$ of the time, the entanglement has decayed completely. Return probabilities between show various levels of entanglement decay.}
    \label{fig:InfBS}
\end{figure}

In standard quantum theory, there is no limit to the number of quantum objects $n$ which can be $n$-partite maximally entangled (i.e., saturate the Coffman-Kundu-Wootters inequality for an $n$-partite state \mbox{\cite{Coffman2000Distributed,Osborne2006CKWInequality})}. However, in IST, there is {(as we saw above)}. Here, we codify this limit, and design experiments using optics/noisy intermediate-scale quantum devices (NISQ devices) to probe it.

For this, we use the $J$-qubit W state \cite{Dur2000Three,Rai2009Possibility},
\begin{equation}\label{EqW}
    \ket{W^J}=\frac{1}{\sqrt{J}}\sum^{J-1}_{i=0}\ket{0}^{\otimes i}\ket{1}\ket{0}^{\otimes (J-1-i)}
\end{equation}
(where $\ket{\psi}^{\otimes J}$ is the tensor product of $\ket{\psi}$ with itself $J$ times). For instance, the W state where $J=3$ is
\begin{equation}
    \ket{W^3}=\frac{\ket{100}+\ket{010}+\ket{001}}{\sqrt{3}}
\end{equation}

The W state is a maximally entangled state of $J$ qubits---and in standard quantum theory, there is no limit to how high $J$ can be.

However, in IST, the finiteness of the $p$-adic metric provides a limit to the number of qudits that can be maximally entangled. {As mentioned above,} for multiple-qubit entanglement, this limit is codified in \cite{Palmer2020Discretization} as a maximum of $\log_2p$ qubits being able to be maximally entangled, in a $p$-adic system where the equatorial great circle of the Bloch sphere consists of $p$ equally-spaced discrete points.

\begin{figure}
    \centering
    \includegraphics[width=\linewidth]{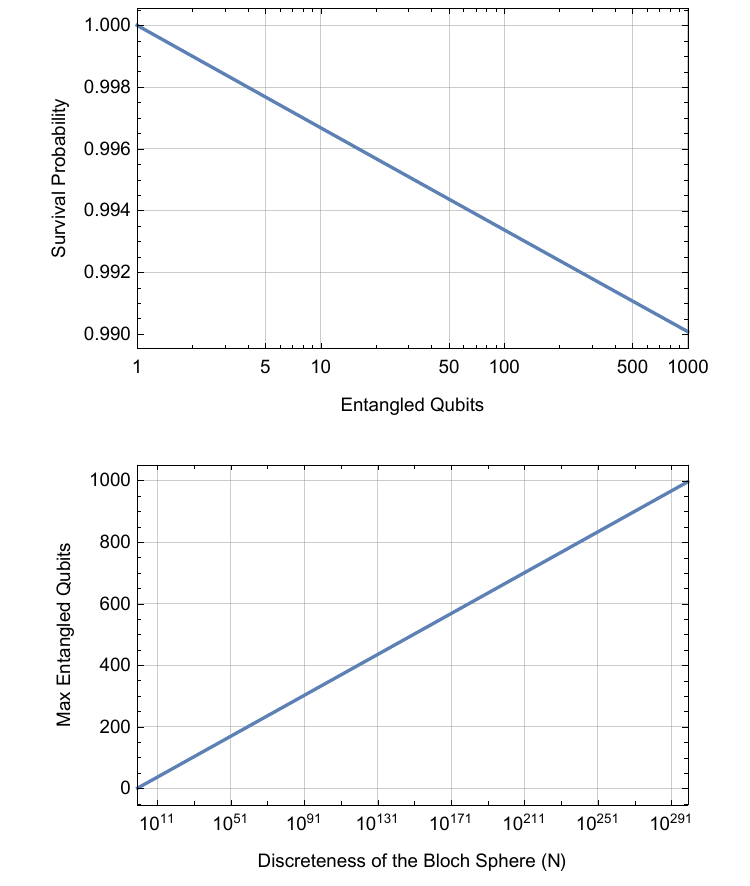}
    \caption{The survival probability of each qubit for a given set of entangled qubits created using the experiment in Fig.~\ref{fig:InfBS}, and the maximum number of entangled qubits that can be created in a version of IST where the $p$-adicity causes the Bloch sphere to be split into $J$ divisions in each angular direction. This shows how this beamsplitter experiment allows us to test this entanglement limit for very high-$p$ versions of IST, due to the comparative lack of loss-induced decoherence on the W state created.}
    \label{fig:Surv}
\end{figure}

A system of maximally-entangled photon-vacuum qubits can be created using a single photon and a number of mirrors and 50:50 beamsplitters, as shown in Fig.~\ref{fig:InfBS}. This naturally forms a W state across $J$ qubits, and, by standard quantum theory, we should potentially be able to extend this set-up to $J\rightarrow\infty$. However, this disagrees with IST, which limits to a maximum of $J=\log_2J$ entangled qubits, where the two orthogonal spherical dimensions of the Bloch sphere ($\theta$ and $\phi$) are each discrete in $J$ divisions. While $p$ is expected to be very large, each qubit will only have been affected by $I=\log_2\log_2J$ beamsplitters, so, for realistic experimental beamsplitter loss of $0.1\%$, the chance of losing a given qubit to decoherence only reaches $1\%$ once the system has entangled over 1000 qubits, which is only possible by IST if $J\geq10^{250}$ (as we show in Fig.~\ref{fig:Surv}). Further, an advantage of the W state is, even if decoherence effectively measures one of the qubits, so long as the result is 0 (the photon isn't in that mode), this collapse leaves the remaining qubits still maximally entangled in the $(J-1)^{th}$ W state.

Even if we obtain this state, we need to prove it is entangled. Gr\"{a}fe et al \cite{Grafe2014WState} and Heilmann et al \cite{Heilmann2015HighOrderW} have done this for 8 and 16 qubit W states respectively, confirming that they generated an entangled W state of that size (assuming they inputted a single photon), and NISQ devices such as Wang et al's integrated silicon photonics chip could be used to do this for a 32-qubit W state \cite{Wang2018Chip}. It is an ongoing problem to specifically discern an entanglement-confirming optical layout for an arbitrarily-large W state, but Lougovski et al give the quantum-information-theoretical groundwork for doing so \cite{Lougovski2009VerifyingWN}. This involves using beamsplitters to shift the optical-path modes to instead each represent one possible permutation of phase combinations for the sub-components (ignoring the global phase of the state). For instance, for the 4-qubit W state, combining beamsplitters after the state creation so as to have each final path act to project on one of the 4 states
\begin{equation}
\begin{split}
    \ket{W_1^4}=(\ket{1000}+\ket{0100}+\ket{0010}+\ket{0001})/2\\
    \ket{W_2^4}=(\ket{1000}-\ket{0100}-\ket{0010}+\ket{0001})/2\\
    \ket{W_3^4}=(\ket{1000}+\ket{0100}-\ket{0010}-\ket{0001})/2\\
    \ket{W_4^4}=(\ket{1000}-\ket{0100}+\ket{0010}-\ket{0001})/2
\end{split}
\end{equation}

Doing this means a consistent detection on just one of the paths over many runs (e.g. the one corresponding to just $\ket{W_1^4}$) indicates a pure entangled state is consistently being created (specifically here the state $\ket{W_1^4}$). Were the entanglement to break, the detections would begin to spread between the targeted state $\ket{W_1^4}$ and the other three states, until, for a maximally mixed state, each detector would click 25$\%$ of the time.

In the same way, for the $I^{th}$ iterate, consisting of $J=2^I$ qubits, using linear optical components one can project the eventual state into one of the $2^I$ phase permutations of $\ket{W^J}$, and so detect with certainty that a pure entangled state of $J$ qubits was created. Interestingly, preparing these states to certify entanglement requires each optical mode to again only interact with $I$ beamsplitters, to allow us to certify $J=2^I$-qubit entanglement, which simply squares the survival probability. This means for 1000 qubits, it becomes $~98\%$ rather than $~99\%$. Given the resilience of the overall state to loss-induced decoherence, and the fact that Lougovski et al show this certification method also allows us to detect any entangled states of fewer than $J$ qubits, this loss probability poses very little issue to our test of IST. Further, despite the loss, the total number of surviving (maximally-entangled) qubits tends to infinity as $I$ tends to infinity, rather than peaking at a certain value.

\section{No Continuous Variables}
A second, related implication of IST is that it permits no continuous quantum variables. As the $p$-adic metric used in IST is necessarily finite-dimensional, the space of states allowed must also be finite. {This can be observed in the bit-string model above through the length limit $p$ per bit-string, and the binary value of each bit---similarly to the finite resolution this allows for the representation of the qubit state on the Bloch sphere, this would only allow a finite coarse-grained value for any continuous variable, rather than an exact specification of a value on a continuum.} Since we can lower bound the number of states allowed as the dimension of the Hilbert space we use (to replicate classical information theory), we can say that, the existence of a qudit of dimension $d$ implies a state space of at least dimension $d$ (e.g. a qubit requires at least two distinct states: 0 and 1; a qutrit requires 3 states: 0, 1 and 2, etc...). Hardy extends this argument, saying that, to satisfy his axioms for quantum theory, between any two pure states in a system, there needs to be a continuous reversible transformation available on a system that goes from one to the other. To allow this, Hardy argues a qudit of dimension $d$ requires a state space of dimension $d^2$ \cite{Hardy2001Axioms}.

This means for continuous variables to exist, given they have an infinite-dimensional Hilbert space \cite{Braunstein2005QIwithCVs}, there must be an infinite number of states allowed. This violates IST. Therefore, in IST, there can be no quantum continuous variables.

In standard quantum physics, many variables are continuous (e.g., position, momentum, electric field strength, and time) \cite{Weedbrook2012Gaussian}. Therefore, for IST to hold true, all of these variables would actually need to be discrete: of finite (but very high) dimension. While some models hold one or another of these variables to be continuous (e.g. space-time in Loop Quantum Gravity \cite{Rovelli1998LoopQGrav,rovelli2011zakopanelecturesloopgravity,Vyas2022LQG} and certain toy models of the Universe \cite{Farrelly2014Discrete}), the idea that all `continuous' variables are actually discrete would be controversial.

\begin{figure}
    \centering
    \includegraphics[width=\linewidth]{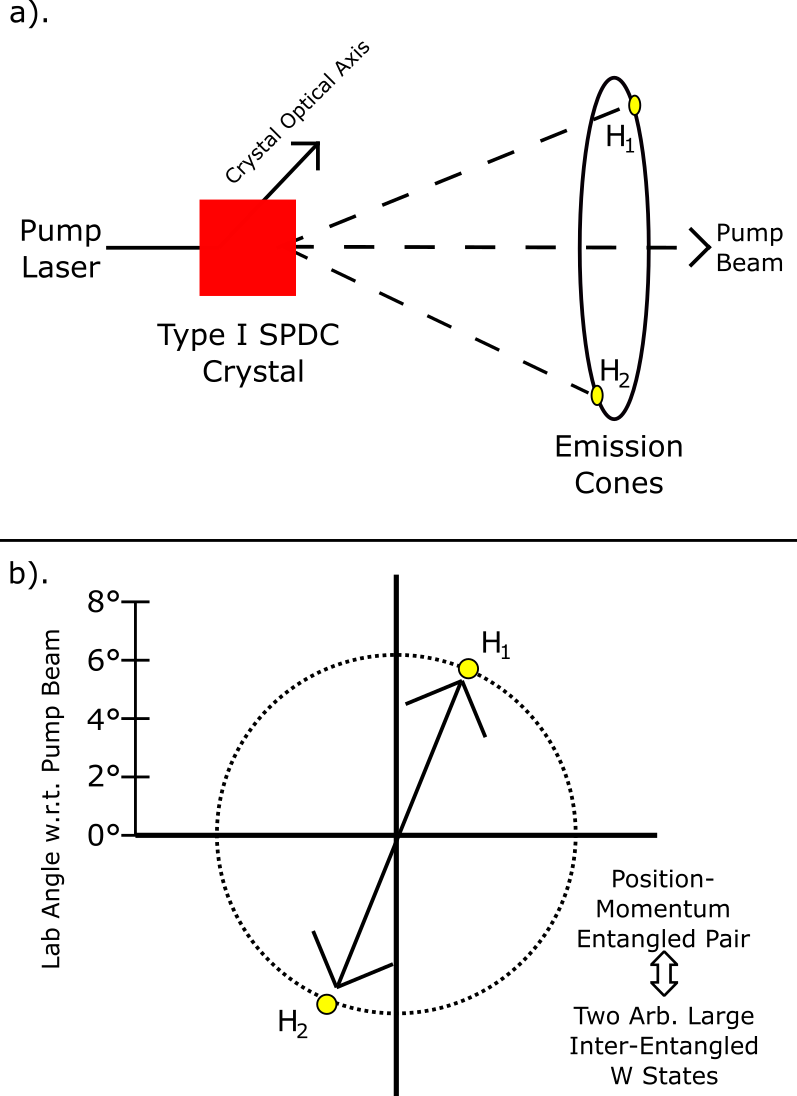}
    \caption{Type I spontaneous parametric down-conversion (SPDC) source for the generation of pairs of position-and-momentum-entangled photons, as given by Rarity and Tapster \cite{Rarity1990Cone}. The generated position of each photon on the cone can be viewed as a W state of arbitrary number of qubits $J$, and so the system of the two photons is a double-W state of $2J$ qubits. This arbitrary number of qubits $J$ can be lower-bounded as the resolution of a circular single-photon position detector array used to detect where on the circle each photon is emitted.
    }
    \label{fig:RTSource}
\end{figure}

{Probing this difference between discrete and continuous treatments of certain variables presents yet another way to test IST---seeing for just how fine-grained a discretisation the experimental effects we would expect (from entangled continuous variables in standard quantum mechanics) still manifest when we instead treat the system as fundamentally discrete and modelled through IST, then probing the system at the level where these predictions diverge.} The W state-based experimental analysis {we give above} can be extended {in this way} by looking at an experiment such as Rarity and Tapster's, where a pair of photons are generated in {what is assumed in standard quantum mechanics to be} a {continuous} cone of possible positions. Here, the angular position of one photon is anti-correlated with the position of the other \cite{Rarity1990Cone}. We show this in Fig.~\ref{fig:RTSource}. Considering just one photon in the cone, this is equivalent to a W state where $J$ is the number of sectors into which you subdivide the cone. Adding a second photon, position-entangled with the first, doubles the number of entangled qubits in the system.

Rarity and Tapster also give a way to prove these photons are entangled: by interfering them to violate a Bell inequality. However, as this is done assuming their position is a continuous variable, we need to adjust it to {identify just what dimension of discrete variable this seemingly-maximal entanglement provably holds for}.

This can be done by making a set of $2J$ apertures on the circumference of the cone, and splitting the ring into two half-circumferences. After this, similarly to what we do in Fig.~\ref{fig:InfBS}, we can iteratively combine adjacent apertures to get position-momentum entanglement between adjacent apertures. Once this projects to equal superpositions across all $J$ apertures on each half-circle, we can record detected position for each half-circle's photon. By comparing the final detected position between upper half-circumference and lower half-circumference, and seeing if they still correlate, we can confirm this double-W$_J$ state.

While the phase between the upper photon and some other discrete division in the upper half will be random, it will be the same as the phase between the lower photon and some discrete division in the lower half. The correlation is always the same, but specific phases at different points on the circumference are not. This is why, using the two photons (and two split half-circles), we can prove the correlations still exist---a similar (continuous-variable) method was used by Rarity and Tapster to provably violate a Bell Inequality.

{According to IST, given the finite resolution available for the representation of both the states of the two photons and the correlations between them, eventually, at a fine enough resolution, we should begin to see the entanglement-induced correlations between the two photons breaking down. As we move to the number of apertures $J$ being of the same order as $p$, IST predicts all entanglement between the two photons should have degraded. However, standard quantum mechanics obviously predicts no aperture-number-dependent degradation of the entanglement-induced correlation. This therefore presents another test we can use to distinguish IST and standard quantum mechanics.}

\section{Gravitational Decoherence}

\begin{figure}
    \centering
    \includegraphics[width=\linewidth]{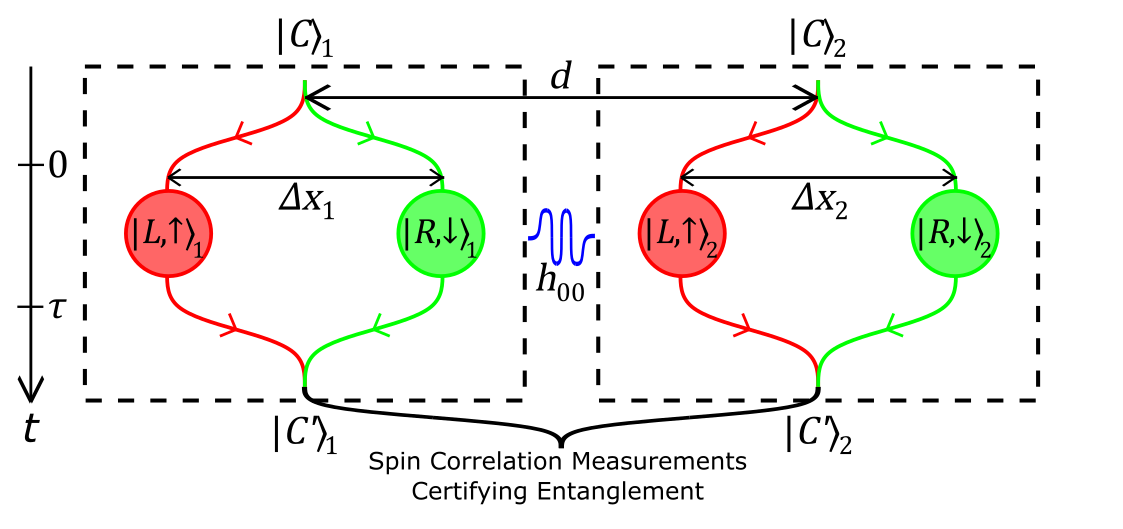}
    \caption{The experiment described by Bose et al \cite{Bose2017EntWitnessQG} and Marletto and Vedral \cite{Marletto2017GravIndEnt}, to test whether gravity can entangle two masses. Two masses, $m_i$ for $i\in\{1,2\}$ are separated from each other by distance $d$. Both are initially in state $\ket{C}_i$, with embedded spin $(\ket{\uparrow}+\ket{\downarrow})/\sqrt{2}$. They are then both admitted into Stern-Gerlach devices, which put them both into the spin-dependent superposition $(\ket{L,\uparrow}_i+\ket{R,\downarrow}_i)/\sqrt{2}$, where $\ket{L}_i$ and $\ket{R}_i$ are separated from each other by distance $\Delta x_i$. They are left in these superpositions for time $\tau$. During this time, if gravity is quantum-coherent, evolution under mutual gravitational attraction $h_{00}$ would entangle the two particles, adding relevant phases to both. After time $\tau$, an inverse Stern-Gerlach device is applied to return each mass to their initial state (potentially modulo the phases applied by $h_{00}$). By applying this process, and measuring spin correlations between the two particles after each run, we can detect if relative phases have been applied to each, and so if gravity is coherent. For IST to hold, gravity must be decoherent, and so cannot entangle two masses. This means IST predicts no alteration of phases will be detected.}
    \label{fig:QGravExp}
\end{figure}

Palmer describes IST as not so much a quantum theory of gravity (like String Theory and Loop Quantum Gravity), but a gravitational theory of the quantum \cite{palmer2016invariantsettheory}. Aside from its determinate nature, nowhere is this more true than in how IST models regimes where gravitational and quantum effects are both present. The paper \textit{Invariant Set Theory} describes the theory as positing no gravitons and so no supersymmetry \cite{palmer2016invariantsettheory} (spin-2 gravitons typically being seen as hinting at supersymmetry \cite{Engelbrecht2022Supersymm}). Instead, the paper holds that gravity is inherently decoherent, turning gravitationally-affected superpositions into maximally mixed states. This paper also claims that effects typically considered signs of either dark matter or dark energy could instead be in some way due to various manifestations of the ``smearing" of energy-momentum on space-times neighbouring our universe $\mathcal{M}_U$ on the invariant fractal set $I_U$ influencing curvature of $\mathcal{M}_U$. It claims this smearing avoids precise singularities in $\mathcal{M}_U$: avoiding singularities being a key goal of many previous attempts to quantise General Relativity.

Palmer suggests an alteration of the Einstein Field Equation (EFE) \cite{misner1973gravitation} based on the presence and effects of possible universes $\mathcal{M}'_U$ on our universe $\mathcal{M}_U$, leading to the EFE instead being
\begin{equation}
\begin{split}
    G_{\mu\nu}&(\mathcal{M}_U)=\\
    &\frac{8\pi G}{c^4}\int_{\mathcal{N}(\mathcal{M}_U)}T_{\mu\nu}(\mathcal{M}'_U)F(\mathcal{M}_U,\mathcal{M}'_U)d\mu
\end{split}
\end{equation}
where $F(\mathcal{M}_U,\mathcal{M}'_U)$ is some propagator to be determined and $d\mu$ is a suitably normalised Haar measure in some neighbourhood $\mathcal{N}(\mathcal{M}_U)$ on $I_U$ \cite{palmer2016invariantsettheory}. Note, in this altered form of the EFE, the cosmological constant $\Lambda$ is set to zero, given Palmer claims the alteration would separately resolve the issue of dark matter and the acceleration of the expansion of the universe.

This gravitational decoherence could be tested by experiments that involve putting heavy objects in spatial superpositions. This would involve allowing them to gravitationally interact, then returning the spatial superposition components back to a single position, then seeing if there are any signs of entanglement between the objects from the resulting interference pattern (see Fig.~\ref{fig:QGravExp}) \cite{Bose2017EntWitnessQG,Marletto2017GravIndEnt,Christodoulou2019Geom,Carney2019TabletopQG}.

In such an experiment, assuming gravity is coherent, the combined state of the two masses initially is
\begin{equation}
    |\Psi_{Init}\rangle_{12}=(\ket{\uparrow}_1 +\ket{\downarrow}_1)(\ket{\uparrow}_2+\ket{\downarrow}_2)\ket{C}_1\ket{C}_2/2
\end{equation}

Passing both masses through a Stern-Gerlach apparatus, this combined state then evolves at $t=0$ to
\begin{equation}
    |\Psi(t=0)\rangle_{12}= (\ket{L,\uparrow}_1+\ket{R,\downarrow}_1)(\ket{L,\uparrow}_2+\ket{R,\downarrow}_2)/2
\end{equation}

After allowing the two masses to gravitationally interact for time $t=\tau$, the overall state has become
\begin{equation}
\begin{split}
     |\Psi&(t=\tau)\rangle_{12}=\frac{e^{i\phi}}{2}\Big(\ket{L,\uparrow}_1(\ket{L,\uparrow}_2
    +e^{i\Delta\phi_{LR}}\ket{R,\downarrow}_2)\\  &+\ket{R,\downarrow}_1(e^{i\Delta\phi_{RL}}\ket{L,\uparrow}_2+\ket{R,\downarrow}_2)\Big)
\end{split}
\end{equation}
where
\begin{equation}
\begin{split}
    &\phi\approx\frac{Gm_1m_2\tau}{\hbar d}\\
    &\phi_{RL}\approx\frac{Gm_1m_2\tau}{\hbar(d-\Delta x)},
    \;\phi_{LR}\approx\frac{Gm_1m_2\tau}{\hbar(d+\Delta x)}\\
    &\Delta\phi_{LR}=\phi_{LR}-\phi,\;\Delta\phi_{RL}=\phi_{RL}-\phi
\end{split}
\end{equation}

After applying the opposite of the initial Stern-Gerlach interaction, the final state is
\begin{equation}
\begin{split}
|\Psi&_{End}\rangle_{12}=\ket{C'}_1\ket{C'}_2=\Big(\ket{\uparrow}_1(\ket{\uparrow}_2+e^{i\Delta\phi_{LR}}\ket{\downarrow}_2)\\
&+\ket{\downarrow}_1(e^{i\Delta\phi_{RL}}\ket{\uparrow}_2+\ket{\downarrow}_2)\Big)\ket{C}_1\ket{C}_2/2
\end{split}
\end{equation}

However, if gravity isn't coherent, there are two possible final states. If gravity doesn't also collapse the state, the final state will be equivalent to the initial one ($\ket{\Psi_{Init}}_{12}=\ket{\Psi_{End}}_{12}$). However, if gravity does collapse the superposition, each particle will be forced into the (spin) maximally mixed state
\begin{equation}
\ket{\Psi_{MM}}_{i}=\ketbra{C}{C}_i(\ketbra{\uparrow}{\uparrow}_i+\ketbra{\downarrow}{\downarrow}_i)/2,\;i\in\{1,2\}
\end{equation}

By measuring spin correlations to estimate the entanglement witness
\begin{equation}
    \mathcal{W}=|\langle\sigma^{(1)}_{x}\otimes \sigma^{(2)}_{z}\rangle-\langle\sigma^{(1)}_{y}\otimes \sigma^{(2)}_{z}\rangle|
\end{equation}
we can distinguish the entangled state from the two other possible final states (if $\mathcal{W}>\mathbb{1}$, the state is entangled), and so see if gravity is coherent. For IST to hold, $\mathcal{W}$ needs to be less than or equal to $\mathbb{1}$.

While such a test may sound challenging to implement, recent experimental work indicates it should be possible within the next few years \cite{Delic2020LevNano,Margalit2021RealisationBose,Gonzalez-Ballestero2021Levitodynamics,Carney2021Atom,Weiss2021Nano}. This test would also have the benefit of either ruling for or ruling out other theories which hold gravity (or some other measure of macroscopicity) to be decoherent: e.g. the Penrose-Di\'osi gravitational collapse model \cite{Diosi1989Reduction,Penrose1996Gravity,Figurato2024DiosiPenrose}, {or other spontaneous collapse models, such as the continuous spontaneous localisation (CSL) model \mbox{\cite{Ghirardi1990CSL,Rijavec2021Decoherence}} or the the Ghirardi-Rimini-Weber (GRW) model \mbox{\cite{Ghirardi1986GRW}}. While these models may seem different to IST, they have one key thing in common with it---none of them hold quantum mechanics to be complete.}

\section{Discussion}

{Unfortunately, all of the tests we discuss above suffer from one key limitation in their ability to discriminate the experimental predictions of Invariant Set Theory from standard quantum mechanics: the fact that $p$ has not yet been bounded in the literature. Note though this is not unusual in the literature, similarly to the difficulties bounding the phenomenological parameters $\lambda$ and $r_C$ for the Ghirardi-Rimini-Weber \mbox{\cite{Ghirardi1986GRW}} and Continuous Spontaneous Localisation \mbox{\cite{Ghirardi1990CSL}} spontaneous collapse models, or the mass-density distribution $R_0$ for the Di\'osi-Penrose gravitational collapse model \mbox{\cite{Diosi1989Reduction,Penrose1996Gravity}}. Identifying the key role played by these free parameters in experimentally differentiating these models from standard quantum mechanics has led to a strong focus in bounding their allowed values from different directions \mbox{\cite{Carlesso2016CollapseBounds,Toros2018Bounds,Vinante2020CollapseParam,Donadi2021Bounds}}, a key aspect of the spontaneous collapse model research programme. We hope, by pointing out that the prospect of finding experimental tests for Invariant Set theory rests heavily on being able to bound $p$, this work will motivate interested researchers to identify similar bounds as has been done for the free parameters of these other models.}

{One question the more reader might have, after reading this paper, is why we have gone to the lengths of considering experimental tests for one specific model, arguably developed enough to only constitute a toy model (rather than a full theory describing and extending beyond quantum mechanics). Our answer is that the meaningfulness of physical models arguably comes from the empirical predictions they make, and specifically the differences in the empirical predictions they make from what is commonly accepted (here, standard quantum mechanics), rather than just from their mathematical structure. We can see this for standard quantum mechanics itself, where the meaningfulness of the completeness/incompleteness debate arguably didn't come from the discussions of Bohr and Einstein \mbox{\cite{einstein1971born,Einstein1947Letter,EPR}}, but from Bell's hypothesis that there would be experimental differences between approaches where quantum mechanics was considered complete (i.e., standard quantum mechanics), and certain approaches where it was considered incomplete (specifically, local hidden variable models respecting statistical independence between measurement settings and hidden variables) \mbox{\cite{Bell1964OnEPR,Bell2004Speakable}}, and Clauser \textit{et al}'s attempts to derive experimental proposals where such a difference was demonstrated \mbox{\cite{CHSH,CH74}}. The fact that the burgeoning field of quantum technologies did not begin at the advent of quantum mechanics, but after standard quantum mechanics had been differentiated from these models is testament to this idea. Elsewhere, we can see this in the rise and fall of String Theory---where interesting mathematical structures attracted many young researchers, the lack of any empirically-testable predictions (and several popular books pointing this out \mbox{\cite{smolin2007trouble,hossenfelder2018lost}}) has left the field slowly wilting away into obscurity. Empirical validity or falsifiability is key for new hypotheses to be meaningful, to attract attention, and ultimately to be usable to develop useful new tools---without this, the hypothesis is arguably just abstract theory or metaphysics. Given (despite not yet being a complete theory) Invariant Set Theory is the archetype for supermeasured theories \mbox{\cite{Hance2022Supermeasured,adlam2023taxonomy}} (those which nominally violate the assumption of statistical independence in Bell's Theorem in a different way to how superdeterministic or retrocausal models are typically assumed to), looking at ways to make IST more physically meaningful is a key research priority for those of us interested in models violating statistical independence. We hope this work, for all its limitations, has succeeded in increasing the meaningfulness of this model.}

To conclude, we have identified points of difference between Invariant Set Theory and standard quantum theory. While these are not fatal to IST, they provide potential avenues to experimentally test the theory, to see whether its deterministic, fractal-attractor-based structure is compatible with observed reality. We described differences between the empirical predictions of IST and standard quantum mechanics. We then proposed experiments, based on current or near-future quantum technologies (e.g. noisy intermediate-scale quantum devices), which would test which set of predictions most closely matches reality. This serves as an example of how these near-term quantum technologies allow us to probe into the foundational mysteries of quantum mechanics. 

\begin{acknowledgments}
JRH acknowledges support from the University of York's EPSRC DTP grant EP/R513386/1, and Hiroshima University's Phoenix Postdoctoral Fellowship for Research programme. JRH and JR acknowledge support from the Quantum Communications Hub funded by the EPSRC grants EP/M013472/1 and EP/T001011/1. TP acknowledges support from a Royal Society Research Professorship. 



\end{acknowledgments}

\bibliographystyle{unsrturl}
\bibliography{ref.bib}

\end{document}